\documentclass[aps,twocolumn,showpacs,superscriptaddress]{revtex4-1}

\usepackage{graphicx}
\usepackage{bm}
\usepackage{amssymb}
\usepackage{amsmath}
\usepackage{dcolumn}
\usepackage{color}
\usepackage{epstopdf}
\usepackage{float}

\begin{document}

\title{Crystal field states of Kondo lattice heavy fermions CeRuSn$_3$ and CeRhSn$_3$}

\author{V. K. Anand}
\altaffiliation{vivekkranand@gmail.com}
\affiliation{ISIS Facility, Rutherford Appleton Laboratory, Chilton, Didcot, Oxon, OX11 0QX, United Kingdom}
\affiliation{\mbox{Helmholtz-Zentrum Berlin f\"{u}r Materialien und Energie GmbH, Hahn-Meitner Platz 1, D-14109 Berlin, Germany}}
\author{D. T. Adroja}
\altaffiliation{devashibhai.adroja@stfc.ac.uk}
\affiliation{ISIS Facility, Rutherford Appleton Laboratory, Chilton, Didcot, Oxon, OX11 0QX, United Kingdom}
\affiliation{\mbox{Highly Correlated Matter Research Group, Physics Department, University of Johannesburg, P.O. Box 524,} Auckland Park 2006, South Africa}
\author{D. Britz}
\affiliation{\mbox{Highly Correlated Matter Research Group, Physics Department, University of Johannesburg, P.O. Box 524,} Auckland Park 2006, South Africa}
\author{A. M. Strydom}
\affiliation {\mbox{Highly Correlated Matter Research Group, Physics Department, University of Johannesburg, P.O. Box 524,} Auckland Park 2006, South Africa}
\affiliation{Max Planck Institute for Chemical Physics of Solids, N\"othnitzerstr.\ 40, 01187 Dresden, Germany.}
\author{J. W. Taylor}
\affiliation{ISIS Facility, Rutherford Appleton Laboratory, Chilton, Didcot, Oxon, OX11 0QX, United Kingdom}
\author{W. Kockelmann}
\affiliation{ISIS Facility, Rutherford Appleton Laboratory, Chilton, Didcot, Oxon, OX11 0QX, United Kingdom}
\date{\today}

\begin{abstract}
Inelastic neutron scattering experiments have been carried out to determine the crystal field states of the Kondo lattice heavy fermions CeRuSn$_{3}$ and CeRhSn$_{3}$. Both the compounds crystallize in LaRuSn$_{3}$-type cubic structure (space group $Pm\bar{3}n$) in which the Ce atoms occupy two distinct crystallographic sites with cubic ($m\bar{3}$) and tetragonal ($\bar{4}m.2$) point symmetries. The INS data of CeRuSn$_{3}$ reveal the presence of a broad excitation centered around 6--8~meV which is accounted by a model based on crystal electric field (CEF) excitations. On the other hand, the INS data of isostructural CeRhSn$_{3}$ reveal three CEF excitations around 7.0, 12.2 and 37.2~meV. The neutron intensity sum rule indicates that the Ce ions at both cubic and tetragonal Ce sites are in Ce$^{3+}$ state in both CeRuSn$_{3}$ and CeRhSn$_{3}$. The CEF level schemes for both the compounds are deduced. We estimate the Kondo temperature $T_{\rm K}= 3.1(2)$~K for CeRuSn$_{3}$ from neutron quasielastic linewidth in excellent agreement with that determined from the scaling of magnetoresistance which gives $T_{\rm K}= 3.2(1)$~K\@.  For CeRhSn$_{3}$ the neutron quasielastic linewidth gives $T_{\rm K}= 4.6$~K\@. For both CeRuSn$_{3}$ and CeRhSn$_{3}$, the ground state of Ce$^{3+}$ turns out to be a quartet for the cubic site and a doublet for the tetragonal site. 

\end{abstract}


\maketitle

\section{INTRODUCTION}

Ce-based heavy fermion systems, whose electronic ground state properties are determined by strongly competing Ruderman-Kittel-Kasuya-Yosida (RKKY) and Kondo interactions, present very interesting physics \cite{Stewart1984,Riseborough, Amato, Lohneysen, Pfleiderer2009, Si2010}. While RKKY interaction tends to establish a long range order, Kondo interaction causes screening of $4f$ moments (below a characteristic Kondo temperature $T_{\rm K}$) leading to a paramagnetic ground state and a quantum critical behavior is realized when the strength of these interactions become comparable, i.e. at the boundary between the magnetically ordered and paramagnetic states. A wide range of intriguing physical properties is seen in the proximity of a quantum critical point (QCP) \cite{Stewart1984,Riseborough, Amato, Lohneysen, Pfleiderer2009, Si2010}. One can achieve QCP by tuning the electronic ground state of antiferromagnetically ordered systems using an external pressure, magnetic field or chemical doping and because of this Ce-compounds are ideal for the study of the physics of quantum critical phenomena. For example, a pressure induced superconductivity is observed in the antiferromagnet CeCoGe$_{3}$ at a critical pressure of 5.5~GPa and a partial substitution of Ge by Si leads to a non-Fermi liquid behavior and quantum criticality in CeCo(Ge$_{1-x}$Si$_x$)$_3$ \cite{Thamizhavel2005, Smidman2013, Settai2007, Eom1998, Continentino2001, Krishnamurthy2002}.

In this paper we focus on two Kondo lattice heavy fermions CeRuSn$_{3}$ and CeRhSn$_{3}$, both of which form in a cubic structure (space group $Pm\bar{3}n$, No.~223) with  LaRuSn$_{3}$ structure as the prototype structure \cite{Eisenmann}. This LaRuSn$_{3}$-type cubic structure consists of two distinct cages formed by the three-dimensional network of trigonal RuSn$_6$ prisms which are occupied by Ce atoms, thus there are two distinct Ce-sites. Earlier investigations of physical properties of CeRuSn$_{3}$ by Fukuhara et al.\ \cite{Fukuhara1989} and Takayanagi et al.\ \cite{Takayanagi1990} revealed a large electronic coefficient $\gamma$ in heat capacity $C_{\rm p}(T)$ and a logarithmic increase ($-\ln T$ behavior) in electrical resistivity $\rho(T)$,  thus characterizing CeRuSn$_{3}$ as a Kondo lattice heavy fermion system. They noticed an absence of coherence effect in resistivity as well as Hall effect, accordingly they suggested the possibility of existence of atomic disorder in CeRuSn$_{3}$ \cite{Fukuhara1989,Takayanagi1990}. They also found a sharp anomaly at 0.5~K in ac susceptibility with an accompanying broad peak in $C_{\rm p}/T$ versus $T$ which they correlated to a magnetic phase transition with an open possibility of antiferromagnetic ordering or spin-glass transition \cite{Takayanagi1990}.  Fukuhara et al.\ \cite{Fukuhara1991} investigated the effect of Sn content on the physical properties of CeRuSn$_{x}$ ($2.85 \leq x \leq 3.15$) and found that the Sn deficiency influences the physical properties. Three phase transitions at 33~K, 4~K and 1.3~K were suggested in CeRuSn$_{2.91}$ with an antiferromagnetic ground state below 1.3~K \cite{Fukuhara1991,Takayanagi1991}. In contrast, no observable change was noticed in the electronic states with Sn content of CeRuSn$_{x}$ for $x=2.85$, 3.0 and 3.15 in core-level photoemission spectroscopy \cite{Ishii1993}.

In our effort to understand the nature of the magnetic transition in CeRuSn$_{3}$ we have investigated this compound using various techniques. Our preliminary results are reported in Ref.~\cite{Anand2015}. Our $C_{\rm p}(T)$ data confirmed the heavy fermion behavior with $\gamma \geq 212(2)$~mJ/mol\,K$^2$  and $\rho(T)$  data confirmed the Kondo lattice feature in CeRuSn$_{3}$ \cite{Anand2015}. We found a broad peak in $C_{\rm p}/T$ versus $T$  near 0.6~K and an anomaly in dc $\chi(T)$ near 0.6~K\@. Further we noticed an irreversibility between the zero-field-cooled and field-cooled $\chi(T)$ data \cite{Anand2015} which is not consistent with the antiferromagnetic model. Furthermore, no long range antiferromagnetic order was inferred from the muon spin relaxation ($\mu$SR) measurement. However, the $\mu$SR data are consistent with glassy spin-dynamics \cite{Anand2015}. Thus our investigations rule out the possibility of previously suggested antiferromagnetic transition in CeRuSn$_{3}$. 

Our investigations of physical properties of CeRhSn$_{3}$ have revealed a complex magnetic ground state and moderate heavy fermion behavior in this compound \cite{Anand2011c}. The $T$ dependence of $\rho$ revealed the Kondo lattice behavior. Both dc $\chi(T)$ and $C_{\rm p}(T)$ exhibit two well pronounced anomalies near 4~K and 1~K\@, while the former seems to be related to a transition to ferrimagnetic ordering, the latter seems to be related to a transition from the ferri- to a ferro-magnetic order below 1~K \cite{Anand2011c}. Interestingly, the ac $\chi(T)$ of CeRhSn$_{3}$ presents an unexplained frequency dependence where the temperature of the 4~K anomaly is found to decrease with increasing frequency. A long-range ordered state below 1~K was inferred from the $\mu$SR study on CeRhSn$_{3}$, however, no such transition was detected at 4~K in the $\mu$SR and powder neutron diffraction studies \cite{Anand2011c}. The isostructural PrRhSn$_{3}$ was found to exhibit a ferromagnetic cluster spin-glass behavior \cite{Anand2012a} in which a frustrated magnetic ground state is believed to result from the dynamic fluctuations of the crystal field levels similar to the case of spin-glass systems PrAu$_{2}$Si$_{2}$ \cite{Goremychkin2008} and PrRuSi$_{3}$ \cite{Anand2011b}. 

Here we extend our investigations on CeRuSn$_{3}$ and CeRhSn$_{3}$ and report the results of inelastic neutron scattering (INS) measurements on these compounds. The INS data are analyzed using a model based on CEF providing information about the crystal field states of Ce in these compounds. A comparitive study of physical properties and INS results of the two compounds are presented.

\section{Experimental}

The polycrystalline samples of CeRuSn$_{3}$ and CeRhSn$_{3}$ as well as their nonmagnetic analogs LaRuSn$_{3}$ and LaRhSn$_{3}$ were prepared by the standard arc melting technique using high purity elements in stoichiometric ratios and subsequent annealing at 900~$^{\circ}$C for 7~days as detailed in Refs.~\cite{Anand2011c,Anand2012a}. The crystal structures and the qualities of the samples were checked by powder x-ray diffraction (XRD) using Cu K$_{\alpha}$ radiation which confirmed the single phase nature and LaRuSn$_{3}$-type cubic crystal structure of all four. Chemical composition was checked by energy dispersive x-ray (EDX) analysis using a scanning electron microscope (SEM) which revealed the desired 1:1:3 stoichiometry. The specific heat was measured by the relaxation method using a physical properties measurement system (PPMS, Quantum Design). Electrical resistivity was measured by the standard four probe method using PPMS.

The room temperature neutron diffraction (ND) experiment was performed on the powdered sample of CeRuSn$_{3}$ using the ROTAX diffractometer at the ISIS facility of Rutherford Appleton Laboratory, Didcot, U.K\@. The inelastic neutron scattering experiment on CeRuSn$_{3}$ was performed with the HET time of flight (TOF) spectrometer at ISIS. The INS experiments on CeRhSn$_{3}$ and LaRhSn$_{3}$ were performed on the MARI TOF spectrometer. The powdered samples of these materials were wrapped in thin Al-foils and mounted inside thin-walled cylindrical Al-cans. Low temperatures down to 4.5~K were obtained by cooling the sample mounts in a top-loading closed cycle refrigerator with He-exchange gas. The INS data were collected for scattering angles between $3^{\circ}$ and $135^{\circ}$ using neutrons with incident energies $E_{i} = 11$~meV, 23~meV and 50~meV for CeRuSn$_{3}$ on HET and $E_{i} = 23$~meV and 50~meV for CeRhSn$_{3}$ and LaRhSn$_{3}$ on MARI.

\section{\label{Sec:crystallography} Crystallography}

\subsection{\label{Sec:CeRuSn3_ND} Powder neutron diffraction study on C\lowercase{e}R\lowercase{u}S\lowercase{n}$_{3}$}

\begin{figure} 
\includegraphics[width=3in]{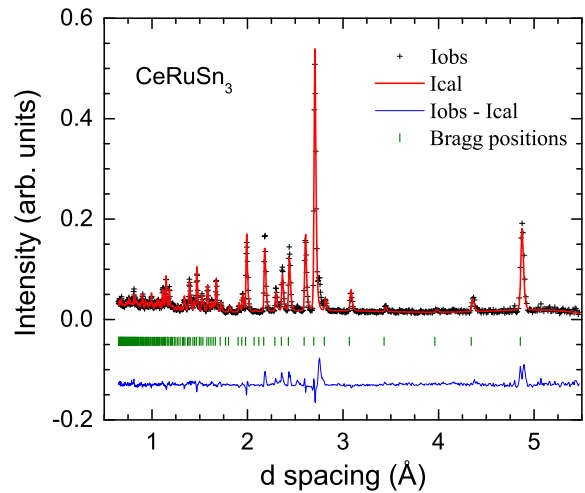}
\caption{\label{fig:ND} (Color online) Powder neutron diffraction pattern of CeRuSn$_{3}$ recorded at room temperature. The solid line through the experimental points is the Rietveld refinement profile calculated for LaRuSn$_{3}$-type cubic (space group $Pm\bar{3}n$) structure. The short vertical bars mark the fitted Bragg peak positions. The lowermost curve represents the difference between the experimental and calculated intensities.}
\end{figure}

The room temperature powder neutron diffraction pattern of CeRuSn$_{3}$ is shown in Fig.~\ref{fig:ND}. The ND data were refined by using the program GSAS \cite{Larson2004}; the structural refinement profile is shown in Fig.~\ref{fig:ND}. The refinement confirmed the LaRuSn$_{3}$-type cubic structure (space group $Pm\bar{3}n$) of CeRuSn$_{3}$ and revealed the single phase nature of the sample. While refining the ND data we checked for the possibility of site mixing of Ru and Sn, however the difference in the scattering lengths of Ru and Sn is too small to produce a noticeable change in the observed intensity for weak site disorder of a few percent. Thus the ND data are of not much help in resolving the possibility of weak atomic disorder suggested in Refs.~\cite{Fukuhara1989,Takayanagi1990}. The refined crystallographic parameters obtained from the least squares refinement of neutron diffraction data are listed in Table~\ref{tab:XRD1}. The structure parameters obtained from the refinements of neutron diffraction and x-ray diffraction data are similar and agree well with the literature values \cite{Eisenmann, Fukuhara1989}. The crystallographic data for CeRhSn$_{3}$ can be found in Ref.~\cite{Anand2011c}.

\begin{table}
\caption{\label{tab:XRD1} Crystallographic parameters obtained from the Rietveld refinement of room temperature powder neutron diffraction data of CeRuSn$_{3}$ using the program GSAS. Profile reliability factor $R_{p} = 10.54\% $ and weighted profile $R$-factor $R_{wp}=9.04\%$.}
\begin{ruledtabular}
\begin{tabular}{lccccc}
\multicolumn{3}{l}{Structure} & \multicolumn{2}{l} {LaRuSn$_{3}$-type cubic}\\
\multicolumn{3}{l}{Space group} & \multicolumn{2}{l} {$Pm\bar{3}n$, No. 223} \\
\multicolumn{3}{l}{Formula units/unit cell}  & \multicolumn{1}{l} {8}\\
\multicolumn{3}{l}{\underline{Lattice parameters}}\\
 \multicolumn{3}{l}{\hspace{0.8 cm}  $a$ (\AA)}        &  \multicolumn{2}{l} {9.7340(4)} \\
 \multicolumn{3}{l}{\hspace{0.8 cm}  $V_{\rm cell}~{\rm (\AA^{3})}$}  &  \multicolumn{2}{l} {922.3(1)}  \\
\multicolumn{3}{l}{\underline{Atomic Coordinates}} \\
 Atom & Wyckoff &	 $x$ 	&	$y$	&	$z$	&  $U_{iso}$ ({\AA}$^{2}$) \\	
 & symbol \\			
	 Ce1 &  2a 	& 0 	& 0          & 0       & 0.072(6)\\
	   Ce2 &  6d 	& 1/4 	&1/2         & 0         & 0.0062(8)\\
	    Ru &   8e 	& 1/4 	& 1/4         & 1/4      & 0.0075(5)\\
     Sn &  24k 	& 0 	& 0.3117(2)   & 0.1571(2)  & 0.0202(7)\\
\end{tabular}
\end{ruledtabular}
\end{table}

\begin{figure} 
\includegraphics[width=\columnwidth, keepaspectratio]{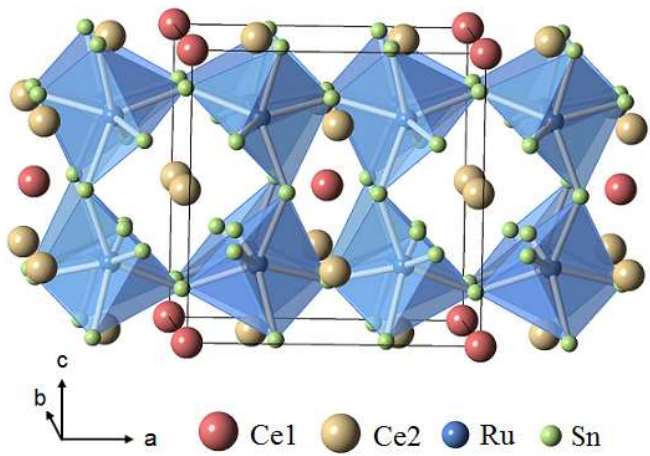}
\caption{\label{fig:structure} (Color online) LaRuSn$_{3}$-type cubic (space group $Pm\bar{3}n$, No. 223) crystal structure of CeRuSn$_{3}$. The spheres represent Ce, Ru and Sn atoms in decreasing order of sizes. Ru atoms are located within the trigonal RuSn$_{6}$ prisms.}
\end{figure}

\subsection{\label{Sec:Crystal} Crystal structure}

The LaRuSn$_{3}$-type cubic structure of CeRuSn$_{3}$ and CeRhSn$_{3}$ is illustrated in Fig.~\ref{fig:structure}. The Ru atoms occupying 8e sites have trigonal prismatic coordination and sit within the tilted RuSn$_{6}$ prisms formed by six Sn atoms. These tilted RuSn$_{6}$ prisms form a three dimensional network and one can observe two different cages which are occupied by Ce atoms. Thus the structure allows two distinct sites for Ce atoms with different near-neighbor environment and coordination numbers (CN).  The Ce atoms sitting at 2a sites referred as Ce1 have a CN of 20 whereas the Ce atoms sitting at 6d sites referred as Ce2 have CN of 16 \cite{Eisenmann,Harmening2010}. Ce1 has a near-neighbor environment of same chemical species consisting of 12 Sn atoms. On the other hand, Ce2 has both Ru and Sn as neighbors. The two Ce sites have different point symmetries -- while Ce1 has a cubic point symmetry ($m\bar{3}$), Ce2 has a tetragonal point symmetry ($\bar{4}m.2$).

\section{\label{Sec:CeRuSn3} C\lowercase{e}R\lowercase{u}S\lowercase{n}$_{3}$}

\subsection{\label{Sec:CeRuSn3INS} Inelastic neutron scattering study}

\begin{figure}
\includegraphics[width=2.5in]{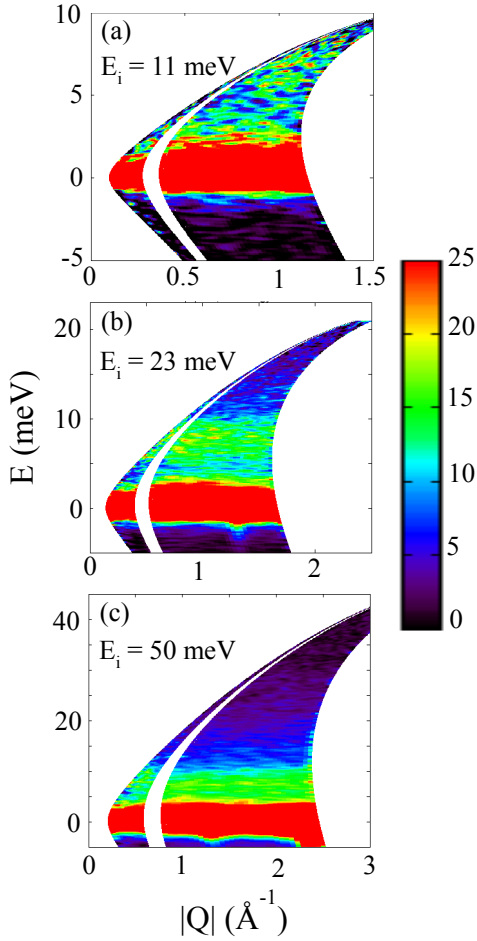}
\caption{\label{fig:INS1} (Color online) Color-coded contour map of inelastic neutron scattering intensity of CeRuSn$_{3}$ measured at 4.5~K with incident energy (a) $E_{i} = 11$~meV, (b) $E_{i} = 23$~meV and (c) $E_{i}= 50$~meV on the HET spectrometer plotted as a function of energy transfer $E$ and wave vector transfer $|Q|$.}
\end{figure}

\begin{figure}
\includegraphics[width=3in]{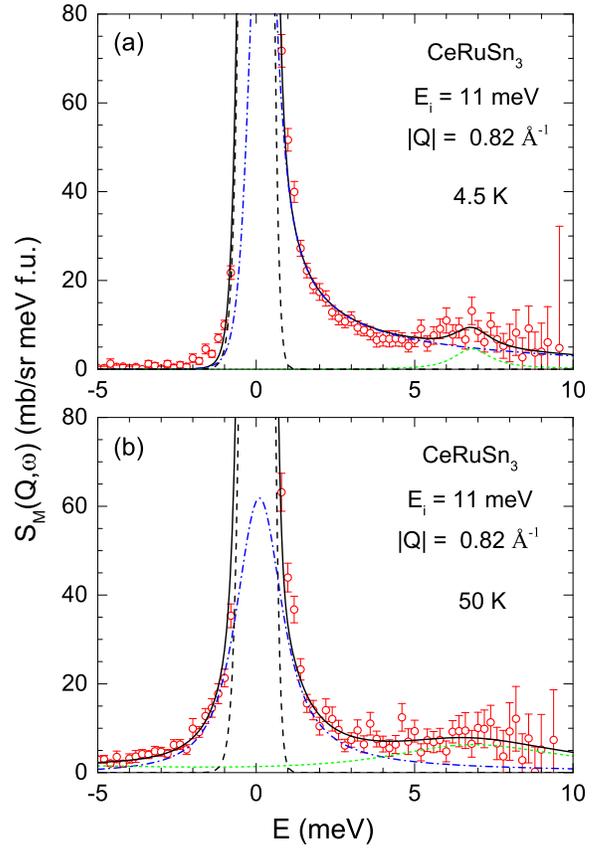}
\caption{\label{fig:INS2} (Color online) $Q$-integrated inelastic magnetic scattering intensity versus energy transfer of CeRuSn$_{3}$ at $|Q|=0.82$~{\AA}$^{-1}$ measured with $E_{i}$ = 11~meV at (a) 4.5~K and (b) 50~K\@. The solid lines are the fits of the data and the dashed and dash-dotted lines are the different components of the fits.}
\end{figure}

Figure~\ref{fig:INS1} shows the INS scattering response, plotted as color-coded intensity maps, at low-$Q$ from CeRuSn$_{3}$ measured with $E_{i} =11$, 23 and 50~meV on HET at 4.5~K\@. There is a clear evidence of magnetic excitations centred around 6--8~meV in the 23 meV and 50 meV data. On the other hand 11~meV data reveal the presence of strong quasi-elastic scattering centred on zero-energy transfer.  The phonon reference compound LaRhSn$_{3}$ did not reveal any strong phonon scattering at low-$Q$ at 4.5~K\@. This indicates that the excitations seen at low-$Q$ in CeRuSn$_{3}$ are due to crystal field excitations of the Ce-ions. It is to be noted that there are two crystallographic sites for Ce atoms in CeRuSn$_{3}$. One Ce site (called Ce1) has cubic point symmetry, $m\bar{3}$, while another Ce site (called Ce2) has tetragonal point symmetry, $\bar{4}m.2$ (Fig.~\ref{fig:structure}). When subject to CEF the six-fold degenerate cubic point symmetry $J=5/2$ state of Ce$^{3+}$ ion will split into a doublet (two fold degenerate, $\Gamma_{7}$) and a quartet (four fold degenerate, $\Gamma_{8}$). Hence for the cubic Ce site we expect one CEF excitation at all temperatures. On the other hand, under the action of CEF,  the tetragonal point symmetry $J=5/2$ state of Ce$^{3+}$ ion will split into three doublets. Thus, if there is no accidental degeneracy, we normally expect two CEF excitations from the ground state of the tetragonal site Ce. However, if the ground state is pure $J_z=\pm1/2$ and excited states are pure $J_z=\pm3/2$  and $J_z=\pm5/2$, we  expect only one CEF excitation from the ground state according to the selection rule for allowed transitions, $\Delta J_z=\pm1$. A similar situation will arise when pure $J_z=\pm5/2$ is a ground state. In both these situations, we expect excited state transitions when temperature is high enough to populate the first excited state.

The crystal field Hamiltonian for the cubic site ($H_{\rm Cubic}$) and tetragonal site ($H_{\rm Tetra}$) Ce atoms are given by
\begin{equation}\label{H-CEF}
\begin{split}
 H_{\rm Cubic} &= B_{4}^{0}[O_{4}^{0} + 5O_{4}^{4}] \\
 H_{\rm Tetra} &= B_{2}^{0}O_{2}^{0} + B_{4}^{0}O_{4}^{0} + B_{4}^{4}O_{4}^{4}
\end{split}
\end{equation}
where $B_{n}^{m}$ are CEF parameters and $O_{n}^{m}$ are Stevens operators. For the cubic site one can determine the value of $B_{4}^{0}= \Delta_{\rm CEF}/360$, where $\Delta_{\rm CEF}$ is the crystal field splitting energy between $\Gamma_{7}$ and $\Gamma_{8}$. However, this gives only magnitude of  $B_{4}^{0}$, not the sign. The sign of $B_{4}^{0}$ (i.e. ground state  $\Gamma_{7}$ for positive $B_{4}^{0}$ or $\Gamma_{8}$ for negative $B_{4}^{0}$) can be determined using the ratio of the INS peak and quasi-elastic linewidth. The calculated  ratio of INS/EQ  is 3.1935 for $\Gamma_{7}$ as a ground state and is 0.6157 for $\Gamma_{8}$ as a ground state.

In order to gain detailed information on the number of CEF excitations and the energy levels scheme, we have analysed the $Q$-integrated one dimensional ($1D$) energy cuts made from the low-$Q$ region (0 to 3~\AA$^{-1}$). The phonon background was subtracted using the scaled high angle (110$^\circ$ to 135$^\circ$) data for 23~meV and 50~meV. The scaling factor was estimated using similar measurements on the nonmagnetic reference compound LaRhSn$_{3}$ on MARI. Figure~\ref{fig:INS2} shows the $1D$ energy cuts at 4.5~K and 50~K from the 11~meV data.  The data were analyzed using the instrument resolution function convoluted with Lorentzian line shape for both quasi-elastic and inelastic excitations.  The population factor and magnetic form factor of Ce$^{3+}$ ion were also included in the fitting  function. It is clear that the data at both the temperatures fit very well to a QE-peak plus one inelastic peak. The value of quasi-elastic linewidth estimated from 11~meV data at 4.5 K is $\Gamma_{\rm QE}(4.5~{\rm K})=0.27(3)$~meV and $\Gamma_{\rm QE}(50~{\rm K})=0.73(9)$. Taking the value of $\Gamma_{\rm QE}(4.5~{\rm K})$ at 4.5 K as $T\rightarrow 0$~K value we have estimated $T_{\rm K}= 3.1(2)$~K using the relation $\Gamma_{\rm QE} = k_{\rm B}T_{\rm K}$, which is in excellent agreement with that estimated from the scaling of magnetoresistance, $T_{\rm K}= 3.2(1)$~K (see Sec.~\ref{Sec:CeRuSn3_MR}). Further the total susceptibility estimated from the QE (0.092~emu/mole) peak and INS (0.0007~emu/mole) peak is 0.092~emu/mole at 4.5~K, which is also in good agreement with the measured susceptibility \cite{Anand2015}.

\begin{figure}
\includegraphics[width=3in]{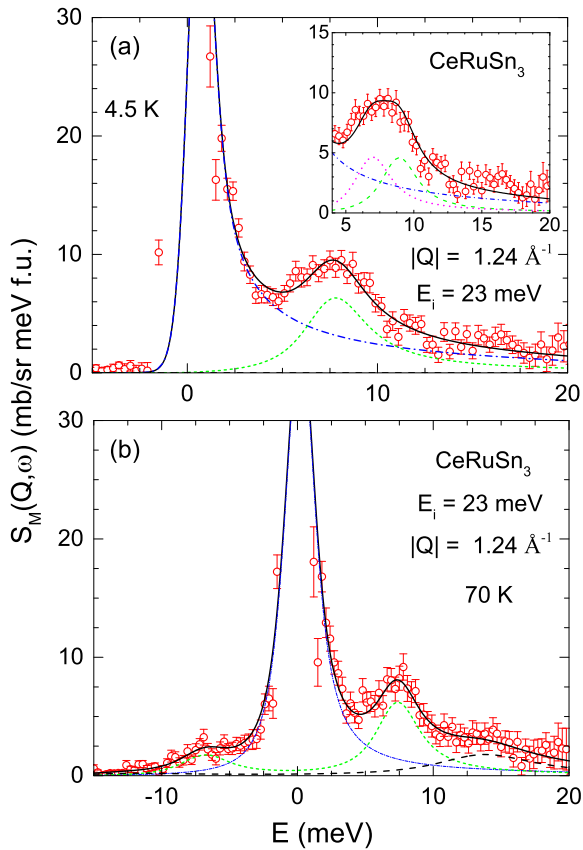}
\caption{\label{fig:INS3} (Color online) $Q$-integrated inelastic magnetic scattering intensity versus energy transfer of CeRuSn$_{3}$ at $|Q|=1.24$~{\AA}$^{-1}$ measured with $E_{i}$ = 23~meV at (a) 4.5~K and (b) 70~K\@. The solid lines are the fits of the data and the dashed and dash-dotted lines are the different components of the fits. The inset in (a) shows the fit of 4.5~K data with two inelastic peaks.}
\end{figure}

\begin{figure}
\includegraphics[width=3in]{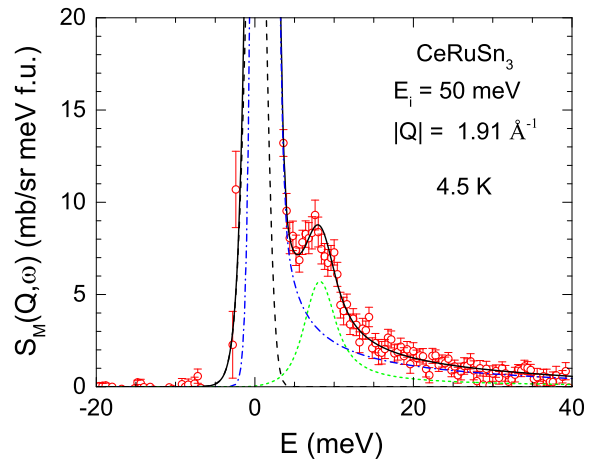}
\caption{\label{fig:INS4} (Color online) $Q$-integrated inelastic magnetic scattering intensity versus energy transfer of CeRuSn$_{3}$ at $|Q|=1.91$~{\AA}$^{-1}$ measured with $E_{i} = 50$~meV at 4.5~K\@. The solid line is the fit of the data and the dashed and dash-dotted lines are the different components of the fit.}
\end{figure}

As there is not much coverage of $Q$-range near 5~meV in data from $E_{i}$ = 11~meV run, we will now discus INS excitations from $E_{i}$ = 23~meV and 50~meV. Figure~\ref{fig:INS3} shows the magnetic scattering estimated from the 23~meV data at 4.5~K and 70~K.  A clear sign of a broad magnetic excitation can be seen at 6--8~meV at 4.5~K. Here we have shown fit to one INS peak, which yields a peak position of 7.4~meV with linewidth of 1.7(2)~meV. It can be seen that although the fit look reasonably good, there is a possibility to have another weak peak below 7.5~meV. We therefore also tried to fit the 4.5~K data with two possible INS peaks, and found that there are two INS excitations at 6.2~meV and 8.2~meV [inset of Fig.~\ref{fig:INS3}(a)]. The 23~meV data at 70~K in [Fig.~\ref{fig:INS3}(b)] could not be fitted with only one INS peak near 7.4~meV and in order to obtain a reasonable fit we had to add another peak near 13~meV. As we did not observe 13~meV peak at 4.5~K in the 23~meV data [Fig.~\ref{fig:INS3}(a)] or in 50~meV data (Fig.~\ref{fig:INS4}), we attribute this peak near 13~meV to an excited state transition from the 6.2~meV (or 8.2~meV) to possible CEF level at 19.2 (or 21.2)~meV. The intensity of the transition from the ground state to 19.2 or 21.2~meV level seems to be extremely weak. 

As we do not have enough information on two separate QE contributions (one from cubic site and another from tetragonal site) and also since there is not enough information on possible three INS excitations (one for cubic site and two for tetragonal site) from the ground state (the CEF excitations from the cubic and tetragonal sites are not resolved), it is not possible to perform a quantitative analysis of the INS data of CeRuSn$_{3}$ using two CEF contributions ($H_{\rm Cubic}$ and $H_{\rm Tetra}$). We therefore use the magnetic contribution to specific heat (see below) in conjunction with the CEF energy level results from INS to determine the ground states of two Ce sites on CeRuSn$_3$. We have checked that the observed magnetic scattering intensity obey the neutron sum rules, $\int S(Q,\omega)/F^2(Q)\,d\omega = 48.6\, \mu_{\rm eff}^2$. The analysis of 11~meV (using quasi-elastic intensity) data and 23~meV (using inelastic intensity) data together gave $\mu_{\rm eff} =  2.6\, \mu_{\rm B}$. This shows that the Ce ions on both the Ce sites are in a 3+ state.

\subsection{\label{Sec:CeRuSn3_HC} Magnetic contribution to heat capacity}

\begin{figure}
\includegraphics[width=3in, keepaspectratio]{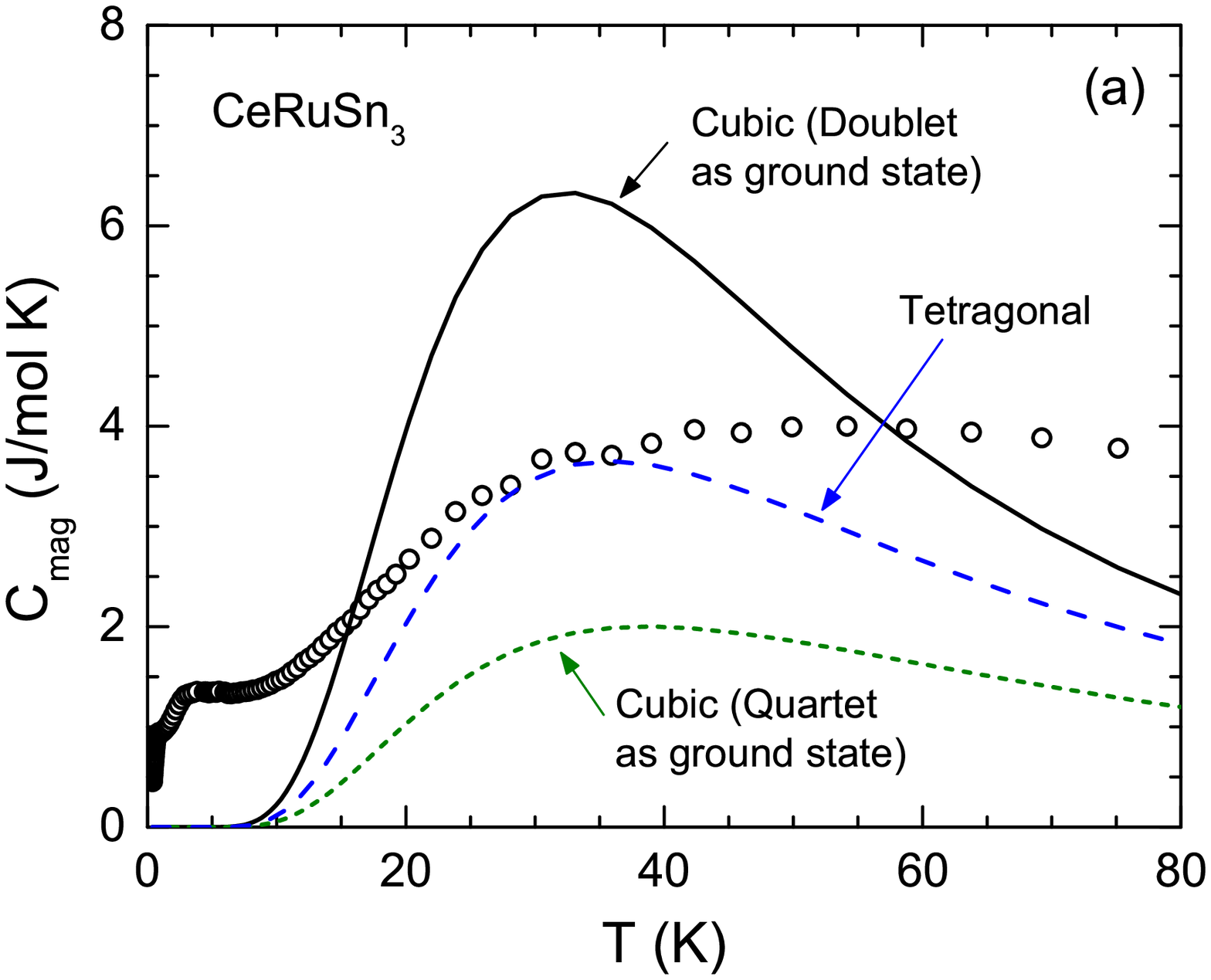}
\includegraphics[width=3in, keepaspectratio]{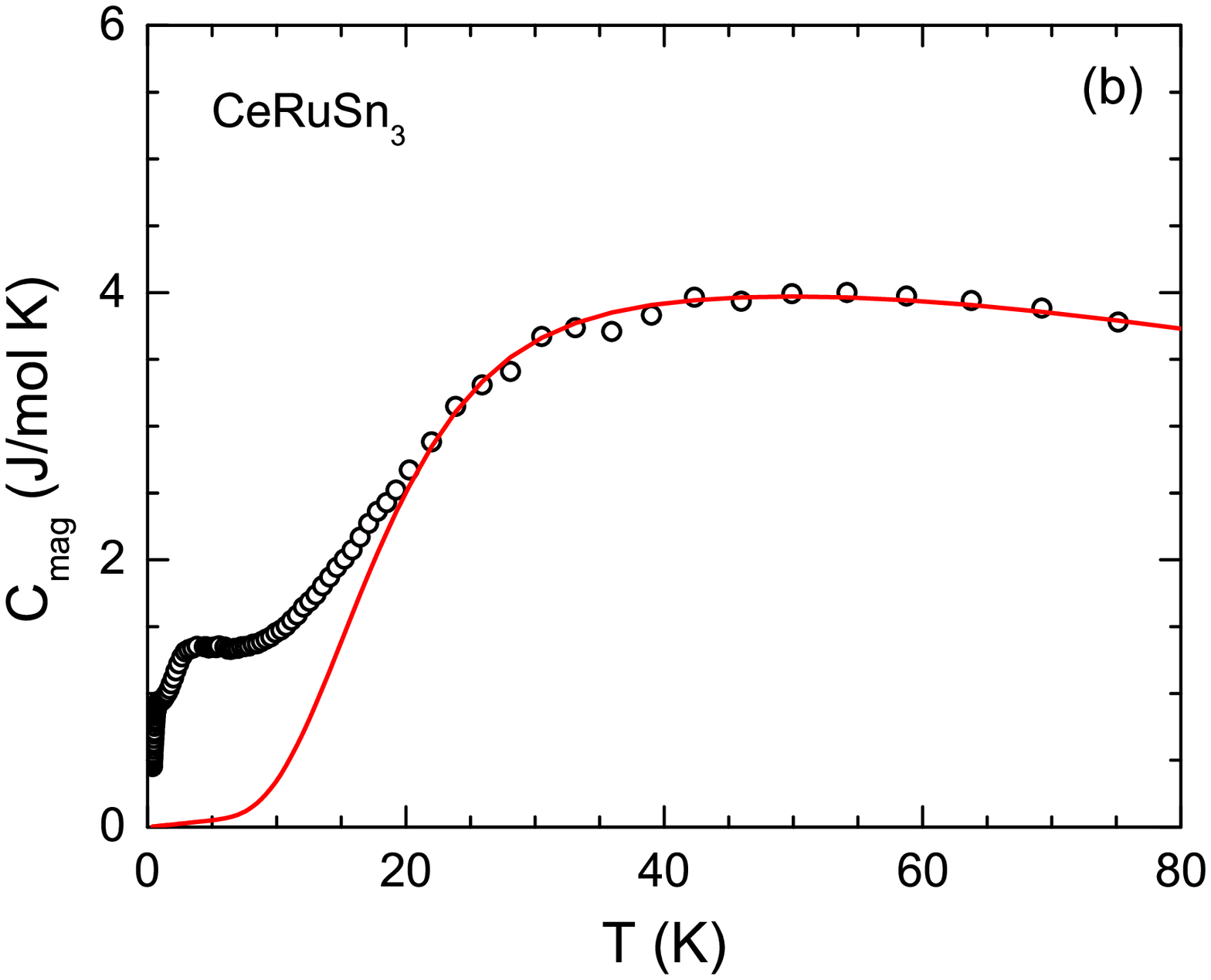}
\caption{\label{fig:HCmag} (Color online) Magnetic contribution to specific heat $C_{\rm mag}$ for CeRuSn$_3$ as a function of temperature $T$\@. The solid and dashed curves represent the crystal electric field (CEF) contributions to specific heat $C_{\rm CEF} (T)$  for (a) two level (splitting energy 7.4~meV) CEF scheme for individual cubic and tetragonal sites, and (b) according to the sum $C_{\rm CEF} =  0.25 \, C_{\rm CEF\,Ce\,Cubic} + 0.75 \, C_{\rm CEF\,Ce\,Tetra} + \gamma T $ taking into account for two Ce sites as discussed in text.}
\end{figure}

In order to check the validity of the CEF level scheme deduced from the analysis of INS data we compare the CEF contribution to specific heat $C_{\rm CEF}(T)$ calculated according to the above CEF scheme with the experimentally determined magnetic contribution to specific heat $C_{\rm mag}(T)$ which is shown in Fig.~\ref{fig:HCmag}. The $C_{\rm mag}(T)$ was obtained by subtracting off the lattice contribution from the $C_{\rm p}(T)$ data of CeRuSn$_3$ assuming the lattice contribution to be roughly equal to that of the nonmagnetic analog LaRuSn$_3$. The $C_{\rm mag}(T)$ exhibits a broad Schottky-type anomaly with a maximum around 40 K which is reproduced by the CEF model. Since there are two Ce sites (2a and 6d) with cubic and tetragonal symmetries, their contribution to specific heat according to their site multiplicities will be 25\% from cubic symmetry Ce and 75\% from tetragonal symmetry Ce, i.e.
\begin{equation}
C_{\rm CEF} =  0.25 \, C_{\rm CEF\,Ce\,Cubic} + 0.75 \, C_{\rm CEF\,Ce\,Tetra},
\label{eq:HC-CEF_sum}
\end{equation}
where $C_{\rm CEF\,Ce\,Cubic}$ represents the contribution from cubic site Ce and Ce $C_{\rm CEF\,Ce\,Tetra}$ that from tetragonal site Ce.

The CEF contribution to specific heat is obtained as \cite{Gopal1966}
\begin{eqnarray}
\label{eq:HC-CEF}
C_{\rm CEF}(T) & &\ =  \left(\frac{R}{T^2}\right) \bigg\{ \sum_{i}g_i {\rm e}^{-\Delta_i/T} \sum_{i}g_i \Delta_i^2 {\rm e}^{-\Delta_i/T}~~~~~~ \nonumber\\
            & &  - \bigg[\sum_{i}g_i \Delta_i {\rm e}^{-\Delta_i/T}\bigg]^2  \bigg\} \times \bigg[\sum_{i}g_i {\rm e}^{-\Delta_i/T}\bigg]^{-2}
\end{eqnarray}
where $g_i$ are the degeneracies of the CEF levels having energies $\Delta_i$. Accordingly for a two-level case one obtains
\begin{equation}
C_{\rm CEF}(T)  = R \left(\frac{\Delta_1}{T}\right)^2 \frac{g_0g_1 {\rm e}^{-\Delta_1/T}}{[g_0 + g_1 {\rm e}^{-\Delta_1/T}]^2},
\label{eq:HC-CEF_TwoLevel}
\end{equation}
and that for the three-level case,
\begin{eqnarray}
C_{\rm CEF}(T) &=& \left(\frac{R}{T^2}\right)\Big\{ g_0g_1 \Delta_1^2 {\rm e}^{-\Delta_1/T} + g_0g_2 \Delta_2^2 {\rm e}^{-\Delta_2/T} \nonumber \\
    & &+~g_1g_2(\Delta_1-\Delta_2)^2 {\rm e}^{-(\Delta_1 + \Delta_2)/T}  \Big\} \nonumber\\
    & & \times~ \Big(g_0 + g_1 {\rm e}^{-\Delta_1/T}+ g_2 {\rm e}^{-\Delta_2/T}\Big)^{-2}
\label{eq:HC-CEF_ThreeLevel}
\end{eqnarray}
where $g_0$, $g_1$ and $g_2$ are the degeneracies of the ground state, first excited state and second excited state, respectively, and $\Delta_1$ and $\Delta_2$ are the energies of the first and second excited states, respectively, with respect to the ground state.

As shown in Sec.~\ref{Sec:CeRuSn3INS}, the INS data (Fig.~\ref{fig:INS3}) could be fitted with one broad peak (7.4~meV) or two unresolved peaks (6.2~meV \& 8.2~meV), therefore we consider both cases for estimating $C_{\rm CEF}(T)$.  First we estimate $C_{\rm CEF}(T)$ for two level CEF scheme (one INS peak) with an energy separation of 7.4~meV which could be the excitation from either of cubic or tetragonal Ce sites. The $C_{\rm CEF}(T)$ obtained [according to Eq.~(\ref{eq:HC-CEF_TwoLevel})] for both cases (cubic Ce site and tetragonal Ce site) are shown in Fig.~\ref{fig:HCmag}(a). As discussed above, for cubic symmetry, the CEF splits the six-fold degenerate state of Ce into a doublet and a quartet, therefore for the cubic case we have considered two possibilities: (a) ground state is a doublet and (b) ground state is quartet. On the other hand, for tetragonal symmetry the CEF splits the six-fold degenerate state of Ce into three doublets, therefore the $C_{\rm CEF}(T)$ was calculated for two doublets separated by 7.4~meV. It is evident from Fig.~\ref{fig:HCmag}(a) that the $C_{\rm mag}(T)$ is not properly represented by two level CEF scheme (one INS peak) in either of the two cases. 

Next we estimate $C_{\rm CEF}(T)$ for the case of two possible INS peaks at 6.2~meV and 8.2~meV according to Eq.~(\ref{eq:HC-CEF_sum}) which takes into account for contributions from both cubic and tetragonal Ce sites. The $C_{\rm CEF}(T)$ estimated according to Eq.~(\ref{eq:HC-CEF_sum}) is shown by the solid red curve in Fig.~\ref{fig:HCmag}(b). The $C_{\rm CEF\,Ce\,Cubic}$ was estimated according to Eq.~(\ref{eq:HC-CEF_TwoLevel}) for the quartet as ground state and doublet as excited state at 8.2~meV, i.e., $g_0 = 4$, $g_1 = 2$ and $\Delta_1=8.2$~meV. The $C_{\rm CEF\,Ce\,Tetra}$ was estimated according to Eq.~(\ref{eq:HC-CEF_ThreeLevel}) for a ground state doublet at 0~meV, the first excited doublet at 6.2~meV and second excited state at 19.2~meV, i.e., for $g_0 = g_1 = g_2 = 2$, $\Delta_1=6.2$~meV and $\Delta_2=19.2$~meV. In addition to contributions from $C_{\rm CEF\,Ce\,Cubic}$ and $C_{\rm CEF\,Ce\,Tetra}$ a $\gamma T $ term corresponding to $\gamma = 10$~mJ/mol\,K$^2$ was added to account for high $\gamma $ of CeRuSn$_3$ that was not properly accounted by the phonon subtaction (equivalent to specific heat of LaRuSn$_3$, which has much lower $\gamma $ than that of CeRuSn$_3$). A very reasonable agreement is found between the experimental $C_{\rm mag}(T)$ data and the calculated $C_{\rm CEF}(T)$ in this case, thus supporting the possibility of two unresolved peaks in INS spectra.

\subsection{\label{Sec:CeRuSn3_MR} Magnetoresistance and Kondo temperature}

\begin{figure}
\centering
\includegraphics[width=3.0in]{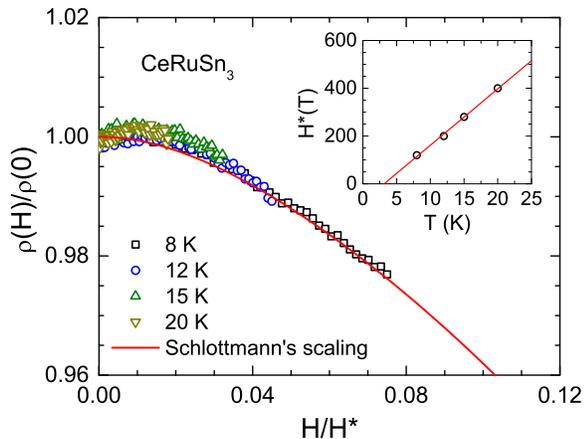}
\caption{\label{fig:Rho_H_scaling} (Color online) Scaled magnetoresistance $\rho(H)/ \rho(0)$ versus reduced field $H/H^*$ of CeRuSn$_{3}$ for various temperatures. Solid curve represents Schlottmann's scaling curve for $J = 1/2$. Inset: Scaling field $H^*$ versus temperature $T$. Solid line is a linear fit of $H^*(T)$ data.}
\end{figure}

The $\rho(H)$ data at different $T$ \cite{Anand2015} allows us to estimate the Kondo temperature within the Schlottmann's description of single-ion Kondo behavior of the Bethe ansatz technique to calculate the magnetoresistance in the Coqblin-Schrieffer model for impurity angular momenta $J \leq$ 5/2 \cite{Schlottmann1983}. Within Schlottmann's model, for a given value of $J$, a universal scaling of MR is found with $H$ and $T$\@. The scaling field $H^*(T)$ is given by \cite{Schlottmann1983}
\begin{equation}\label{Eq:rhoH_Kondo1}
    H^*(T)  = H^*(0) + \frac{k_{\rm B}T}{g \mu},
\end{equation}
where $k_{\rm B}$ is Boltzmann's constant, $g$ is the Lande factor, $\mu$ is the moment of Kondo ion and the Kondo field $ H^*(0)$ is related to Kondo temperature $T_{K}$ by
\begin{equation}\label{Eq:rhoH_Kondo2}
   H^*(0)  = \frac{k_{\rm B}}{g \mu} T_{\rm K}.
\end{equation}
 The normalized resistivity $\rho(H)/\rho(0)$ of CeRuSn$_{3}$ as a function of the reduced field $H(T)/H^*(T)$ is shown in Fig.~\ref{fig:Rho_H_scaling}. The solid red curve in Fig.~\ref{fig:Rho_H_scaling} represents the Schlottmann's scaling curve for a CEF-split Kramers doublet ground state with an effective $J=1/2$ (or $S=1/2$) of the Ce$^{3+}$ ion. A very reasonable agreement is observed between the theoretical curve and the $H$ and $T$ scaled experimental MR data. A plot of $H^*$ versus $T$ is shown in the inset of Fig.~\ref{fig:Rho_H_scaling}. A fit to the $H^*(T)$ by Eq.~(\ref{Eq:rhoH_Kondo1}) shown by solid line in the inset of Fig.~\ref{fig:Rho_H_scaling} gives the Kondo temperature $T_{\rm K} = 3.2(1)$~K, in excellent agreement with $T_{\rm K} = 3.1(2)$~K estimated from the neutron quasi-elastic linewidth above. 

It should be noted though that because of the presence of two Ce sites, where for cubic symmetry site Ce ions neutron scattering and specific heat data suggest a quartet to be the ground state, one would not expect the model based on effective pseudo spin $J = 1/2$ to describe the MR data very precisely. Nevertheless, as the majority (three fourths) of Ce ions belong to tetragonal symmetry site having a doublet as ground state, the MR data are reasonably approximated by the model of Schlottmann's scaling for $J = 1/2$. Furthermore from Fig.~\ref{fig:Rho_H_scaling} we note that at low fields the MR data show a positive curvature which is different from the case of CeRhSn$_{3}$ where no such positive curvature is observed \cite{Anand2011c}, though they both have negative curvature at higher fields. This difference possibly reflects that magnetic exchange in CeRuSn$_{3}$ is of antiferromagnetic nature whereas CeRhSn$_{3}$ has a ferri-/ferro-magnetic exchange interaction.

Another estimate of $T_{\rm K}$ follows from the Weiss temperature using the relation \cite{Gruner1974} $T_{\rm K} \approx |\theta_{\rm p}|/4.5 $, which for $\theta_{\rm p} = -17.2(3)$~K of CeRuSn$_{3}$ \cite{Anand2015} gives $T_{\rm K} \approx  3.8$~K\@ which is very close to the above estimated values of $T_{\rm K}$. A broad hump near 3~K in the $C_{\rm mag}(T)$ data (Fig.~\ref{fig:HCmag}) could be associated with $T_{\rm K}$.

A rough estimate of high-$T$ Kondo temperature $T_{\rm K}^{\rm h}$ can be obtained from the value of $\gamma$ within the Coqblin-Schrieffer model using the relation \cite{Yamamoto, Rajan, Tsvelick}
 \begin{equation}\label{Eq:Kondo_temph}
  T_{\rm K}^{\rm h}  =   \frac{W J \pi R}{3 \gamma}
\end{equation}
applicable to a dense Kondo system, where $W = 0.1026 \times 4 \pi$ is the Wilson number, $R$ is the molar gas constant and $J = 5/2$ for Ce$^{3+}$. This for $\gamma = 212(2)$~mJ/mol\,K$^2$ \cite{Anand2015} gives $T_{\rm K}^{\rm h} = 132$~K\@. This value of $T_{\rm K}^{\rm h}$ is close to the value $T_{\rm K}^{\rm h} = 120$~K obtained by Fukuhara et al.\ \cite{Fukuhara1991} using the Hamman-Fisher law. However, our value of $T_{\rm K} \approx 3.2$~K is much lower than the value $T_{\rm K} = 20$~K they obtained using the Hamman-Fisher law \cite{Fukuhara1991}.

\begin{figure}
\includegraphics[width=2.5in]{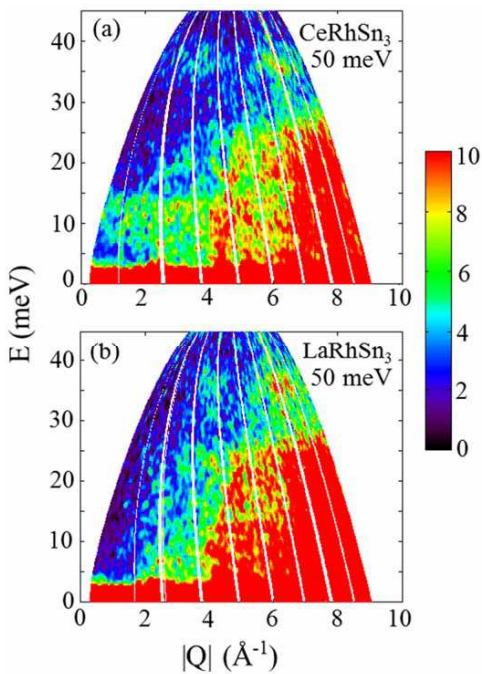}
\caption{\label{fig:CeRhSn3_INS1} (Color online) Color-coded contour map of inelastic neutron scattering intensity of (a) CeRhSn$_{3}$ and (b) LaRhSn$_{3}$ measured at 4.5~K with incident energy $E_{i} = 50$~meV on the MARI spectrometer plotted as a function of energy transfer $E$ and wave vector transfer $|Q|$.}
\end{figure}

\section{\label{Sec:CeRhSn3} C\lowercase{e}R\lowercase{h}S\lowercase{n}$_{3}$}
\subsection{Inelastic neutron scattering study}

\begin{figure}
\includegraphics[width=3in]{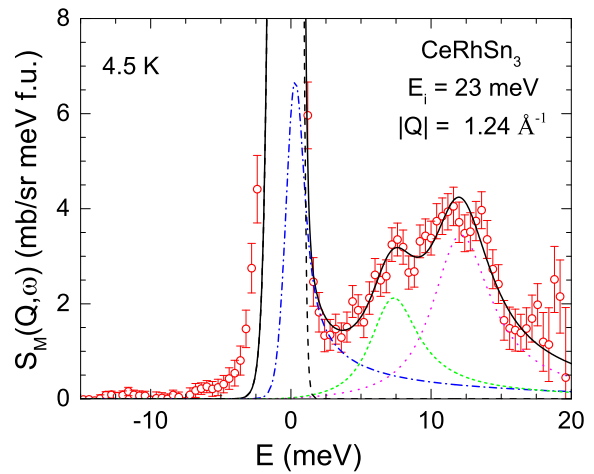}
\caption{\label{fig:CeRhSn3_INS2} (Color online) $Q$-integrated inelastic magnetic scattering intensity versus energy transfer of CeRhSn$_{3}$ at $|Q|=1.24$~{\AA}$^{-1}$ for $E_{i}$ = 23~meV measured  at 4.5~K on MARI. The solid line is the fit of the data and the dashed and dash-dotted lines are the different components of the fits.}
\end{figure}

\begin{figure}
\includegraphics[width=3in]{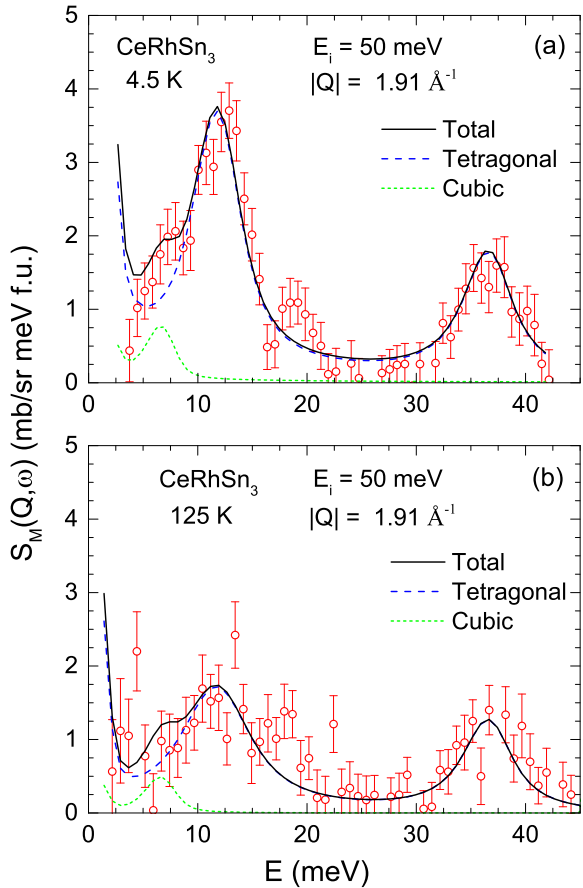}
\caption{\label{fig:CeRhSn3_INS3} (Color online) $Q$-integrated inelastic magnetic scattering intensity versus energy transfer of CeRhSn$_{3}$ at $|Q|=1.91$~{\AA}$^{-1}$ measured with $E_{i}$ = 50~meV at (a) 4.5~K and (b) 125~K\@. The solid lines are the fits of the data by CEF model accounting for two Ce sites together with the contributions for cubic and tetragonal sites shown by dashed curves. Note: The apparent weak peak near 19~meV is an artifact due to phonon.} 
\end{figure}

The color coded intensity maps showing INS scattering responses from CeRhSn$_{3}$ and LaRhSn$_{3}$ measured with $E_{i} =50$~meV at $T= 4.5$~K are shown in Fig.~\ref{fig:CeRhSn3_INS1}. While at low-$Q$ no excitation is seen for LaRhSn$_{3}$, three magnetic excitations near 7.0, 12.2 and 37.2~meV are clearly evidenced for CeRhSn$_{3}$. The magnetic scattering $S_{\rm M}(Q,\omega)$ to CeRhSn$_{3}$ INS response was obtained after subtracting the phonon background using the similar INS measurement on the nonmagnetic reference compound LaRhSn$_{3}$. The $S_{\rm M}(Q,\omega) = S(Q,\omega)_{\rm CeRhSn_3} - \alpha\,S(Q,\omega)_{\rm LaRhSn_3}$ with $ \alpha = 0.75$, the ratio of neutron scattering cross sections of CeRhSn$_3$ and LaRhSn$_{3}$. The $1D$ energy cuts at 4.5~K from the 23~meV and 50~meV data in the low-$Q$ region (0 to 3~\AA$^{-1}$) are shown in Figs.~\ref{fig:CeRhSn3_INS2} and \ref{fig:CeRhSn3_INS3} for $E_{i}$ = 23~meV  and $E_{i}$ = 50~meV, respectively. The three magnetic excitations at 7.0, 12.2 and 37.2~meV are very clear in the $1D$ cut of CeRhSn$_{3}$ INS data in Figs.~\ref{fig:CeRhSn3_INS2}  and \ref{fig:CeRhSn3_INS3}. We attribute these excitations to the crystal field excitations from the two Ce sites, one for the cubic Ce site (most probably near 7.0~meV) and another two from the tetragonal Ce site. As the raw data shown in Fig.~\ref{fig:CeRhSn3_INS1} do not show any magnetic excitation near 19 meV, the weak peak near 19 meV in Fig.~\ref{fig:CeRhSn3_INS3} seems to be a spurious artifact consequent to the phonon subtraction procedure or background scattering of unknown origin (see Appendix, Fig.~\ref{fig:CeRhSn3_INS_Raw}). We also measured the INS response at 125~K for $E_{i}$ = 50~meV, which did not show any additional excitation towards the transition among excited states.

As in the case of CeRuSn$_{3}$ the INS data were analysed using the Lorentzian shape for both quasi-elastic and inelastic excitations, the fitting is shown in Fig.~\ref{fig:CeRhSn3_INS2}. The fit of 23~meV INS data yielded the quasi-elastic linewidth at 4.5 K,  $\Gamma_{\rm QE}(4.5~{\rm K})=0.4(1)$~meV. This in turn gives $T_{\rm K}= 4.6 \pm 1.2$~K which is close to but somewhat higher than the previous estimate of $T_{\rm K}= 2.4$~K that was obtained from the scaling of magnetoresistance \cite{Anand2011c}. For a precise estimate of $T_{\rm K}$ from $\Gamma_{\rm QE}$ the INS data should have been collected at low incident energy. However, as we do not have INS data at energy lower than 23~meV we used the $\Gamma_{\rm QE}$ from the INS data measured at 23~meV to estimate $T_{\rm K}$, therefore this estimate of $T_{\rm K}$ may not be very accurate. The $\theta_{\rm p} = -12.2$~K of CeRhSn$_{3}$ \cite{Anand2011c} gives $T_{\rm K} \approx 2.7$~K\@. 

Further we analyzed the INS data of CeRhSn$_{3}$ using CEF contributions given in Eq.~(\ref{H-CEF}), which takes into account of two crystallographic sites. Our simultaneous analysis of the 50~meV INS data at 4.5~K and 125~K yielded the best fit for the CEF parameters listed in Table~\ref{tab:CEF}. The fit of the 50~meV INS data at 4.5~K and 125~K are shown in Fig.~\ref{fig:CeRhSn3_INS3}. The analysis shows that the ground state for the cubic site is a quartet and that for the tetragonal site is a doublet, which agrees well with the ground state of CeRuSn$_{3}$ inferred from the combined analysis of heat capacity and INS data in Sec.~\ref{Sec:CeRuSn3}. The CEF wave functions obtained for the cubic site Ce are:
 \begin{equation}\label{Eq:Ce_Cubic}
\begin{split}
 \Psi_1  = &\    (0.4082) |\pm \frac{3}{2}\rangle +  (0.9129) |\mp \frac{5}{2}\rangle\\ 
            &\ \& ~|\pm \frac{1}{2}\rangle \\
\Psi_2  = &\    (0.9129) |\pm \frac{3}{2}\rangle -  (0.4082) |\mp \frac{5}{2}\rangle
\end{split}
\end{equation}
with quartet $\Psi_1$ as ground state and doublet $\Psi_2$ as first excited state at 6.84~meV. The CEF wave functions for the tetragonal site Ce are:
 \begin{equation}\label{Eq:Ce_Cubic}
\begin{split}
\Psi_1 & =   (0.9995) |\pm \frac{3}{2}\rangle +  (0.0302) |\mp \frac{5}{2}\rangle \\
\Psi_2  & =    |\pm \frac{1}{2}\rangle \\
\Psi_3  & =   (-0.0302) |\pm \frac{3}{2}\rangle +  (0.9995) |\mp \frac{5}{2}\rangle.
 \end{split}
\end{equation}
The energy eigenvalues for the three doublets $\Psi_1$ (ground state), $\Psi_2$ (first excited state) and $\Psi_3$ (second excited state) are 0, 12.04 and 36.97~meV, respectively. 

\begin{figure}
\includegraphics[width=3in, keepaspectratio]{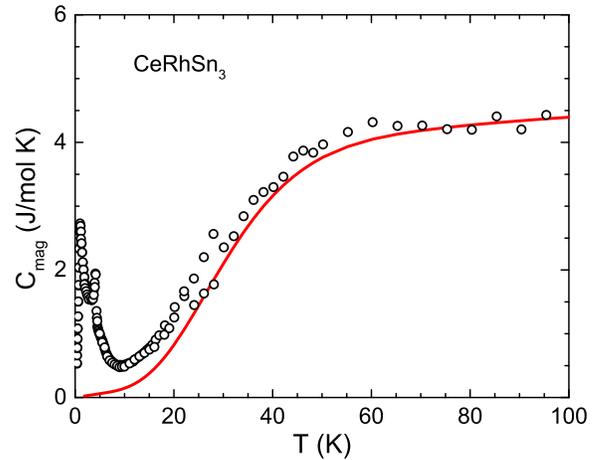}
\caption{\label{fig:HCmag2} (Color online) Magnetic contribution to specific heat $C_{\rm mag}$ for CeRhSn$_3$ as a function of temperature $T$\@. The crystal electric field (CEF) contributions to specific heat according to the CEF level scheme obtained from the analysis of the inelastic neutron scattering data corresponding to the sum $C_{\rm CEF} =  0.25 \, C_{\rm CEF\,Ce\,Cubic} + 0.75 \, C_{\rm CEF\,Ce\,Tetra} + \gamma T $.}
\end{figure}

\subsection{Magnetic contribution to heat capacity}

The $C_{\rm mag}(T)$ data of CeRhSn$_3$ are shown in Fig.~\ref{fig:HCmag2} and the CEF contribution to specific heat $C_{\rm CEF}(T)$ estimated for the above obtained CEF level scheme according to Eq.~(\ref{eq:HC-CEF_sum}) is shown by the solid red curve.  An additional $\gamma T $ [over that of LaRhSn$_3$ that was subtracted off to obtain $C_{\rm mag}(T)$] corresponding to $\gamma = 10$~mJ/mol\,K$^2$ was added to calculated $C_{\rm CEF}(T)$ to compare with $C_{\rm mag}(T)$ data. A very reasonable agreement between the experimental $C_{\rm mag}(T)$ data and the calculated $C_{\rm CEF}(T)$ supports the obtained CEF level scheme.

\begin{table}
\caption{\label{tab:CEF} Crystal field parameters $B_{n}^{m}$ and splitting energies $\Delta_i$ of excited states (with respect to ground state, $\Delta_0 \equiv 0$) obtained from the analysis of the inelastic neutron scattering data of CeRhSn$_{3}$.}
\begin{ruledtabular}
\begin{tabular}{lcc}
     & Cubic & Tetragonal\\
$B_2^0$ (meV) &  & $+1.62(1)$  \\
$B_4^0$ (meV) & -0.019 & $+0.072(8)$ \\
$B_4^4$ (meV) &  & $-0.042(1)$ \\

$\Delta_1$ (meV) & 6.84 & 12.04\\
$\Delta_2$ (meV) &  & 36.97 \\

\end{tabular}
\end{ruledtabular}
\end{table}

\section{Discussion}

It is interesting to note that despite having a very similar Kondo temperature, the compounds CeRuSn$_{3}$ and CeRhSn$_{3}$ exhibit very different physical properties. While CeRuSn$_{3}$ does not show long range magnetic order down to 84~mK \cite{Anand2015} and is situated close to an antiferromagnetic QCP, CeRhSn$_{3}$ exhibits complex magnetic order below 4~K with dominant ferri-/ferro-magnetic interaction \cite{Anand2011c}. Apparently in the case of CeRuSn$_{3}$ the Kondo interaction wins over the RKKY interaction and suppresses the long range ordering of Ce$^{3+}$ moments, whereas in the case of CeRhSn$_{3}$, RKKY interaction dominates over the Kondo interaction leading to a long range ordering. The difference between the magnetic behaviors of CeRuSn$_{3}$ and CeRhSn$_{3}$ seems to be related to the presence of transition metals Ru ($4d^75s^1$) and Rh ($4d^85s^1$), having different number of $4d$ electrons. The extra $d$-electron in Rh moves CeRhSn$_3$ towards more localized nature than CeRuSn$_3$. Such effect of electron doping and shift towards more localized limit has also been observed in Ce(Ru$_{1-x}$Rh$_x$)$_2$Al$_{10}$ \cite{Kobayashi2013, Kimura2015} and Ce(Fe$_{1-x}$Rh$_x$)$_2$Al$_{10}$ \cite{Tanida2015}. 

The strength of hybridization between the conduction electron and the $4f$ electron (c-$f$ hybridization) is transition metal dependent. The Kondo temperature depends on c-$f$ hybridization strength $\Gamma$ and density of states at Fermi level ${\cal D}(E_{\rm F})$, $T_{\rm K} \sim \exp(-|\epsilon_f|/ \Gamma^2{\cal D}(E_{\rm F}))$ where $\epsilon_f$ is the binding energy of the $4f$ level. The $\gamma$ values of about 100~mJ/mol\,K$^{2}$ for CeRhSn$_{3}$ \cite{Anand2011c} and 212~mJ/mol\,K$^{2}$ for CeRuSn$_{3}$ \cite{Anand2015} clearly reflect that CeRuSn$_{3}$ has much larger ${\cal D}(E_{\rm F})$than that of CeRhSn$_{3}$. The similar values of $T_{\rm K}$ despite the different ${\cal D}(E_{\rm F})$ can thus be naively considered to be an indication for the two compounds to have different c-$f$ hybridization strength. Their different inelastic neutron scattering responses and hence crystal field excitations possibly reflect the difference in hybridization strength associated with different numbers of $4d$-electrons. 

The ternary compounds CeRhSi$_{3}$ and CeRuSi$_{3}$ both having BaNiSn$_{3}$-type noncentrosymmetric tetragonal (space group $I4\, mm$) structure also show different physical properties because of the different degree of c-$f$ hybridization by transition metals Rh and Ru. CeRhSi$_{3}$ is found to order antiferromagnetically below 1.6~K and exhibit pressure induced superconductivity at a critical pressure of about 1.2~GPa \cite{Muro1998,Kimura2005}. In contrast, with a dominant Kondo interaction CeRuSi$_{3}$ remains paramagnetic \cite{Kawai2008}. The INS study revealed evidence for hybridization gap in CeRuSi$_{3}$ \cite{Smidman2015}. A similar magnetically ordered versus paramagnetic ground state for Rh and Ru has been observed in the case of ThCr$_{2}$Si$_{2}$-type tetragonal (space group $I4/mmm$) structure compounds CeRh$_{2}$Si$_{2}$ and CeRu$_{2}$Si$_{2}$. While CeRh$_{2}$Si$_{2}$ orders antiferromagnetically below 36~K \cite{Graf1998,Kawarazaki2000}, no evidence of long range ordering is seen down to 170~$\mu$K in CeRu$_{2}$Si$_{2}$ \cite{Takahashi2003}. An extremely small static electronic moment of the order of $10^{-3}~\mu_{\rm B }$ was reported from $\mu$SR measurement on CeRu$_{2}$Si$_{2}$ \cite{Amato1994}. The INS study shows different CEF splitting energies in these compounds: 0, 32, 33~meV for CeRu$_{2}$Si$_{2}$ and 0, 30, 52~meV for CeRh$_{2}$Si$_{2}$ \cite{Willers2012}. A recent angle resolved photo electron emission study on CeRh$_{2}$Si$_{2}$ suggests a significant mixing of Ce $4f^1$ and $4f^0$ states even in the antiferromagnetic state demonstrating the importance of hybridization effects \cite{Patil2016}.

\section{Conclusions}

We have performed inelastic neutron scattering experiments on two Kondo lattice heavy fermion systems CeRuSn$_{3}$ and CeRhSn$_{3}$. An estimate of Kondo temperature is made from the neutron quasi-elastic linewidth which gave a value of $T_{\rm K} = 3.1(2)$~K for CeRuSn$_{3}$ in excellent agreement with the estimate of $T_{\rm K} = 3.2(1)$~K from the scaling of magnetoresistance data. A trivalent state of Ce ions for both the cubic and tetragonal Ce sites is inferred from the neutron intensity sum rule. In contrast to the expected three CEF excitations (one from cubic and two from tetragonal Ce cites), the INS data of CeRuSn$_{3}$ reveal a broad excitation near 6--8~meV. The analysis of INS data using a CEF model shows the possibility of two unresolved CEF excitations near 6.2 and 8.2~meV with an overall CEF splitting of about 19.2~meV which is supported by the analysis of $C_{\rm mag}(T)$ data based on the CEF model. The INS data of CeRhSn$_{3}$ on the other hand clearly exhibit three CEF excitations near 7.0, 12.2 and 37.2~meV. From the neutron quasi-elastic linewidth analysis the $T_{\rm K}$ for CeRhSn$_{3}$ is estimated to be $T_{\rm K} \approx 4.6$~K\@. The ground state of Ce$^{3+}$ in both the compounds is found to be quartet for the Ce$^{3+}$ ions occupying the cubic site and doublet for the Ce$^{3+}$ ions occupying the tetragonal site. The crystal field parameters and ground state wave functions for both the cubic and tetragonal sites of Ce$^{3+}$ were determined for CeRhSn$_{3}$.

It appears that the transition metal (Rh and Ru with different number of $4d$ electrons) plays a decisive role in controlling the c-$f$ hybridization strength which in turn governs the physical properties, manifesting different crystal field excitations and hence different inelastic neutron scattering responses for the two compounds. Further investigations of hybridization strength would be enlightening for the understanding of the role of hybridization in determining the nature of electronic ground state of these compounds.

\acknowledgements
DTA and VKA acknowledge financial assistance from CMPC-STFC grant number CMPC-09108. DTA also thanks JSPS for funding support to visit Hiroshima University and Prof. Takabatake for kind hospitality during the visit. AMS thanks the SA-NRF (93549) and the URC of UJ for financial assistance. DB thanks the URC of UJ for a GES Scholarship which enabled him to participate in this work as part of his PhD studies.

\appendix*
\section{Comparison of scattering from CeRhSn$_{3}$ and LaRhSn$_{3}$}

\begin{figure}
\includegraphics[width=2.6in]{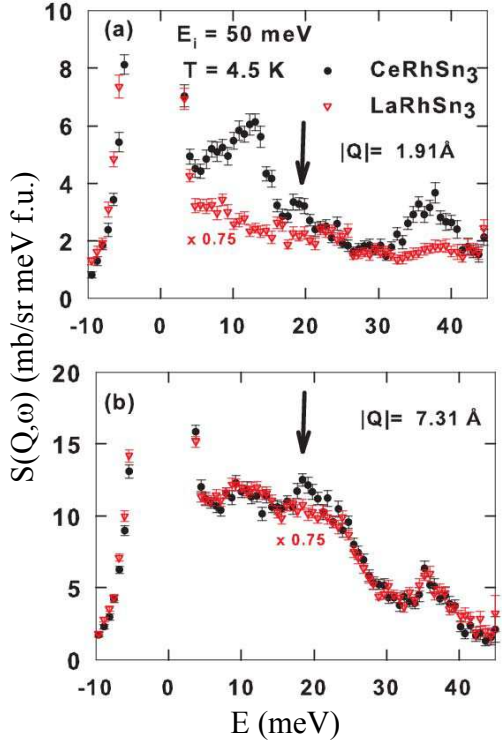}
\caption{\label{fig:CeRhSn3_INS_Raw} (Color online) $Q$-integrated inelastic scattering intensity versus energy transfer of CeRhSn$_{3}$ and LaRhSn$_{3}$ at (a) low $|Q|=1.91$~{\AA}$^{-1}$ and (b) high $|Q|=7.31$~{\AA}$^{-1}$ for $E_{i}$ = 50~meV measured  at 4.5~K on MARI. The LaRhSn$_{3}$ data have been scaled by a factor $\alpha = 0.75$. The arrows show the spurious peak of nonmagnetic origin near 19~meV.}
\end{figure}

A comparison of the $Q$-integrated INS scattering from CeRhSn$_{3}$ and LaRhSn$_{3}$ at 4.5~K at low $Q$ (1.91~{\AA}$^{-1}$) and high $Q$ (7.31~{\AA}$^{-1}$) is shown in Fig.~\ref{fig:CeRhSn3_INS_Raw} for $E_{i}$ = 50~meV. In order to estimate the phonon contribution to INS data of CeRhSn$_3$ the INS data of LaRhSn$_{3}$ have been scaled by a factor $\alpha = 0.75$ equivalent to the ratio of neutron scattering cross sections of CeRhSn$_3$ and LaRhSn$_{3}$. The difference plot $S_{\rm M}(Q,\omega) = S(Q,\omega)_{\rm CeRhSn_3} - S(Q,\omega)_{\rm LaRhSn_3}$ is presented in Fig.~\ref{fig:CeRhSn3_INS3}(a) showing the magnetic scattering from CeRhSn$_3$ at low $Q$. A comparison of low-$Q$ and high-$Q$ $S(Q,\omega)$ of ${\rm CeRhSn_3}$ shows that the weak peak near 19~meV marked with arrow, that is present even at high-$Q$, is not of magnetic origin. The intensity of excitations of magnetic origin decreases with increasing $Q$ and becomes negligibly small at high enough $Q$. In contrast we see that the 19~meV peak is present at both low and high $Q$. On the other hand the three magnetic excitations near 7.0, 12.2 and 37.2~meV disappear at high $Q$. This suggests that the 19~meV peak does not originate from CEF excitations, could be due to phonon or some unknown background scattering. Thus the apparent weak peak near 19~meV in $S_{\rm M}(Q,\omega)$ in Fig.~\ref{fig:CeRhSn3_INS3} is not due to CEF excitation.

\end{document}